# Fan-out and Fan-in properties of superconducting neuromorphic circuits

*M. L. Schneider* [1,a], *K. Segall* [2,a]
[1] National Institute of Standards and Technology Boulder, CO
[2] Colgate University, Hamilton, NY

**Abstract**
Neuromorphic computing has the potential to further the success of software-based artificial neural networks (ANNs) by designing hardware from a different perspective. Current research in neuromorphic hardware targets dramatic improvements to ANN performance by increasing energy efficiency, speed of operation, and even seeks to extend the utility of ANNs by natively adding functionality such as spiking operation. One promising neuromorphic hardware platform is based on superconductive electronics, which has the potential to incorporate all of these advantages at the device level in addition to offering the potential of near lossless communications both within the neuromorphic circuits as well as between disparate superconductive chips. Here we explore one of the fundamental brain-inspired architecture components, the fan-in and fan-out as realized in superconductive circuits based on Josephson junctions. From our calculations and WRSPICE simulations we find that the fan-out should be limited only by junction count and circuit size limitations, and we demonstrate results in simulation at a level of 1-to-10,000, similar to that of the human brain. We find that fan-in has more limitations, but a fan-in level on the order of a few 100-to-1 should be achievable based on current technology. We discuss our findings and the critical parameters that set the limits on fan-in and fan-out in the context of superconductive neuromorphic circuits.

**I Introduction**

In recent years there has been an increase in the study of neuromorphic circuits made from superconducting electronics.[1 2 3 4 5 6 7 8 9 10 11 12] The single flux quantum (SFQ) pulse that is released from a Josephson junction when the applied current briefly exceeds its critical current has a lot in common with the action potential in neurons that is released when the membrane potential exceeds its threshold. Superconducting circuits based on Josephson junctions are being developed to generate pulses, send pulses down output lines, transmit variable feed-forward signals, and sum the signals from different sites in a network; these circuits represent the electrical equivalent of biological somas, axons, synapses and dendrites, respectively. As a result, it is possible to conceive of building spiking neural networks (SNNs) completely from superconducting electronics. SNNs are used for pattern recognition in spatio-temporal data sets and are considered by some to have superior information-processing capabilities to the basic artificial neural networks (ANNs) used in "deep learning."[13] They are also good analogs of the human brain and are used to study its computational capabilities.[14] SNNs made from superconducting electronics have the potential to be both faster and more energy-efficient[15 16] than those made from silicon circuits.

In superconducting digital circuits, the presence or absence of an SFQ pulse during a clock cycle represents the 1 or 0 of digital logic. As a result a large variety of circuits that manipulate and control SFQ pulses are available including flip-flops, logic gates, shift registers, transmission lines, splitters, mergers and comparators.[17] In moving to neuromorphic circuits, much of the effort has been to simply redesign and repurpose these circuits to perform the necessary functions in a neuromorphic[18] environment. Fan-in is

a) M. L. Schneider and K. Segall contributed equally to this work.

the number of inputs that a circuit element can accept, whereas fan-out refers to the number of loads that can be connected to a circuit element's output. One of the weaknesses of SFQ logic is its poor fan-in and fan-out, typically only 2 or 3 at most.[19] Given the extraordinary fan-out of the human brain (~ 10,000), this is a huge concern for the utility of SFQ based neuromorphic circuits.

In this paper we explore the problem of fan-out and fan-in in superconducting neuromorphic circuits. Although there are many similarities to the fan-in and fan-out properties of superconducting digital circuits, there are also important differences, most notably the analog nature of the synapse. We make some basic assumptions about a possible system architecture and show both flux- and current-based methods of fan-in and fan-out. We use both circuit simulations and calculations to show how these approaches might scale and what factors might limit them in pushing to larger networks. With modest assumptions about fabrication parameters like inductances and critical currents, we show how a fan-out and fan-in of order 100 is possible. Although this falls short of the human brain, it brings the technology in line with silicon neuromorphic computing, without loss of the advantages of low power and high speed.

The paper is organized as follows. In Section II we give an overview of the problem, assuming a basic system architecture. In Section III we show two possible ways of fanning out, including splitter trees and a current fan-out. In Section IV we show two methods of fan-in, one with flux and one with current. In Section V we look at the limiting factors for each of these as one scales to larger networks. Finally, in Section VI we conclude and suggest future work.

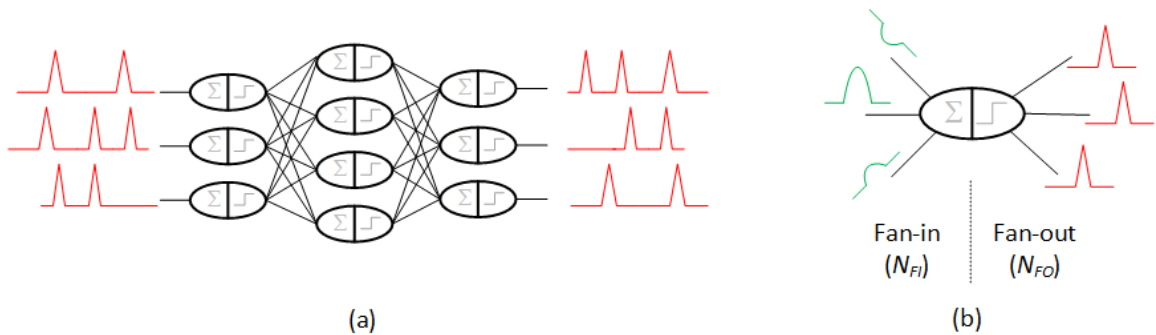

Figure 1 Schematic of the fan-in and fan-out of a spiking neural network (SNN). 1a) Schematic of a SNN, with the neurons indicated as ovals and the synapses indicated as connecting lines. Neural spike patterns in time are the inputs and outputs of the system. The neurons are divided into two halves: the input half, which sums the inputs, and the output half, which generates a spike when the threshold is exceeded. The weighting occurs along each connecting line and is not shown explicitly. 1b) The number of signals that are summed at the input form the fan-in ($N_{FI}$), while the number of output lines that a spike is driven down forms the fan-out ($N_{FO}$). The input signals are of varying weight, both positive and negative, while the output signals are all the same amplitude and shape.

**II Overview**

In this section we carefully define the problem of fan-in and fan-out, first in generalized spiking neural networks and then, by assuming a simplified architecture, in a superconducting electronics implementation. Since there is as of yet no agreed-upon architecture for a superconducting neural network, some of our results will change as different architectures are proposed and tested. However, we believe that many of the ideas developed here are general enough that they will translate into future experiments regardless of architecture.

Figure 1 shows a general schematic diagram of a spiking neural network (SNN). Indicated are spiking neurons, synapses, and propagation lines that represent axons and dendrites. Neurons are broken into two parts: the input, where signals from many synapses are summed together, and the output, where a spike is emitted when the sum of the inputs exceeds a threshold. The number of signals summed together on the input constitutes the fan-in, while the number of output lines that a spike is propagated down is the fan-out. The fan-in and fan-out are about the same (on average) for a given network, since most signals start in one neuron and end in another, with the exception of the inputs and outputs to the system.

One important distinction needs to be made about the signals from neurons and synapses. The signals that flow out of neurons (fan-out), called the spikes or "action potentials" in a biological setting, are essentially *digital*: they all have about the same amplitude and shape, which are fixed by the ion channel dynamics inside the soma. For this reason, the SFQ pulses of superconducting digital logic are ideal to serve as action potentials. Meanwhile, the signals that flow out of synapses into neurons (fan-in), the so-called "post-synaptic potentials," are *analog* in nature: although they typically have a similar shape, their amplitude depends on the strength of the synapse and they can be either positive (excitatory) or negative (inhibitory).[20] In many implementations of a spiking neural network, synaptic signals are represented digitally; for example, IBM's True North used a 4-bit digital number to represent the amplitude of the synaptic current.[21] However, this represents a quantity which is still inherently analog and the bit depth that is required to capture this analog behavior is the topic of active investigation.[22]

These ideas make the role of fan-in and fan-out circuitry clearer. For fan-in, we need to sum the analog signals from different synapses coming into the soma. For fan-out, we need to take the output signal of the soma and generate identical copies, each to be sent down a separate line to a synapse.

In our superconducting implementation, we assume a simple two-junction neuron similar to the Josephson junction neuron (JJ neuron) already proposed.[1] Although a single junction will generate a spike when driven above threshold, for larger input currents an unwanted offset will develop; this is avoided in a two-junction neuron. The two junctions also give the neuron more biological realism, as each junction can be likened to an ion channel, with the resulting reproduction of several biologically realistic neuron behaviors.[1] The input to the neuron is either a current or a flux, and the threshold is defined when an SFQ pulse is produced from a given input. This can be tuned by adjusting circuit parameters such as a bias current.

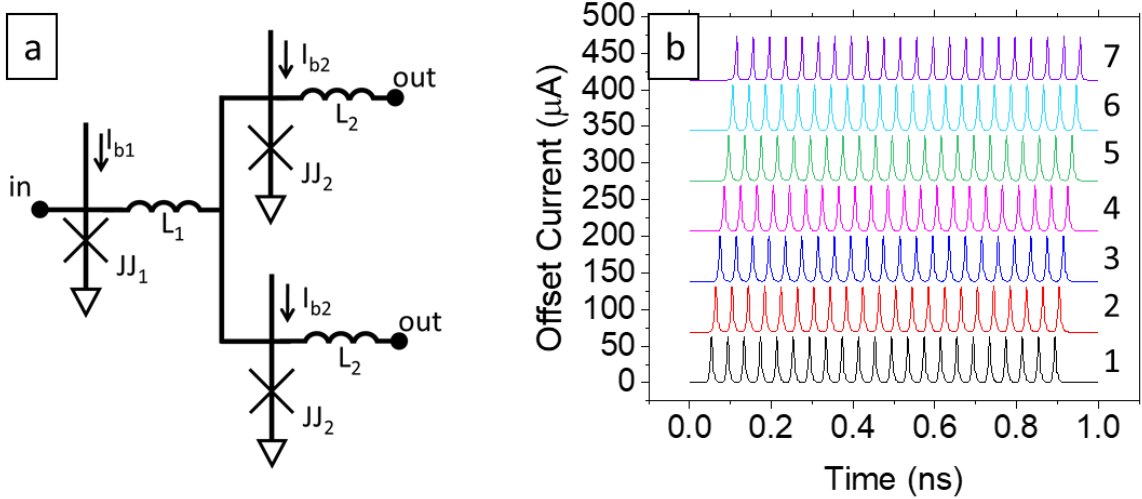

*Figure 2a) Schematic of binary splitter. 2b) Output current pulses sampled at a random $L_2$ inductor at each of the layers in a 1-to-128 fan-out circuit of nested binary splitters with layers labeled to the right of the traces. The data shown in 2b are from a simulation with the following circuit parameter: $JJ_1$ has Ic = 70 µA, $JJ_2$ has Ic = 50 µA, $L_1$ = 15 pH, $L_2$ = 20 pH, $I_{b1}$ = 50 µA, and $I_{b2}$ = 35 µA.*

### III Fan-out

In this section we will discuss two potential methods, flux based and current based, for fan-out that can be used in superconducting neuromorphic circuits. Because the fan-out in these circuits is effectively a digital operation, one can start with a fan-out circuit that is a series of nested splitters like those used typically in digital SFQ circuits.[17] Figure 2a shows the schematic of the binary splitter circuit, which will be nested in layers to achieve a larger fan-out. The input to the first layer is an SFQ pulse coming from a "pre-synaptic" JJ neuron and the output at the final splitter layer will be transmitted to a synapse. In the intermediate layers, the input will be from the previous splitter layer and the output will go to the subsequent splitter layer. With a network of such splitters one can fan-out to a given power of 2 with an overhead of $3N_{FO}-3$ junctions, where $N_{FO}$ is the final fan-out number. It is worth noting that the junction biases can be run in series meaning that only two lines would be needed for the fan-out circuit.

Figure 2b shows the current pulse at each layer of a binary splitter, as labeled to the right of the trace, that has been nested seven times to achieve a fan-out of 1-to-128. Note that the output is nearly identical for each trace aside from the slight time delay, as desired. The output was taken at a random location within a given layer and corresponds to the current through the inductor after division ($L_2$). In this example we feed an initial DC to SFQ converter with a 25 GHz triangle wave, its output is fed into the input of the first splitter layer at the node marked "in" in Fig. 2a. The latency through this fan-out is roughly 6 ps. In this example we chose the critical current of $JJ_1$ to be $I_{c1}$ = 70 µA with $L_1$ = 15 pH and $JJ_2$ to have $I_{c2}$ = 50 µA with $L_2$ = 20 pH. We also chose the junctions to be overdamped, quantified by $\beta_c$ = 0.35, where $\beta_c = \frac{2\pi I_c R_n^2 C_j}{\phi_0}$ is the McCumber parameter, and set the bias current to 70% of the respective $I_c$ values. This method provides a straightforward means of copying the signal flowing out of a JJ neuron to an arbitrarily large number of synapses.

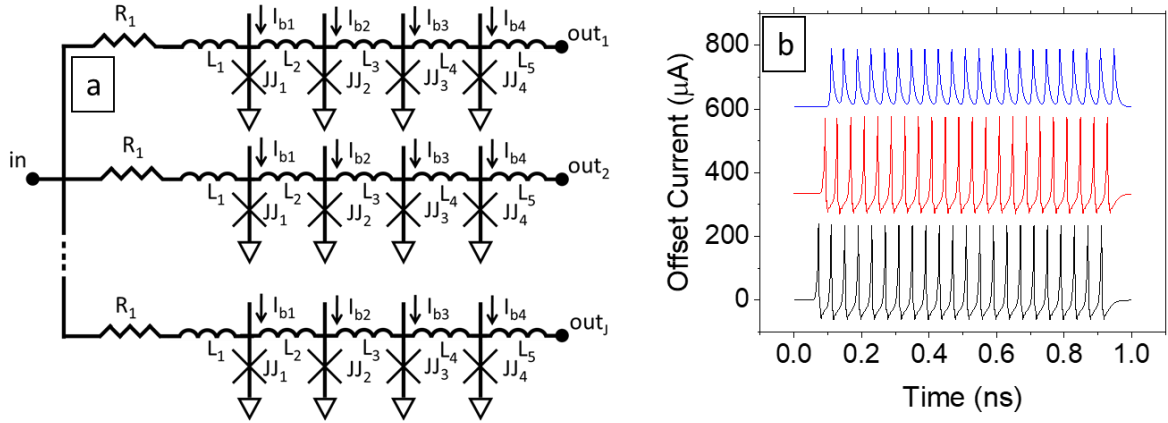

*Figure 3a) Schematic of a 1 to N current fanout circuit.  3b) Output current pulses from a random output at each of the layers in a 1-to-1000 fan-out circuit of by 10 current splitters (pulses are offset for clarity with the first stage at the bottom and the last stage at the top).  The data shown in 3b are from a simulation with the following circuit parameter; $R_1$ = 5 Ohms, $JJ_1$ has $I_c$ = 25 µA, $JJ_2$ has $I_c$ = 50 µA, $JJ_3$ has $I_c$ = 110 µA, $JJ_4$ has $I_c$ = 250 µA, $L_1$ = 1 pH, $L_2$ = 40 pH, $L_3$ = 20 pH, $L_4$ = 10 pH, $L_5$ = 2.5 pH, $I_{b1}$ = 22.5 µA, $I_{b2}$ = 45 µA, $I_{b3}$ = 99 µA, and $I_{b4}$ = 225 µA.*

In addition to the flux-based fan-out shown in Fig. 2, one can also use a current based fan-out as shown in Fig. 3a.  In this fan-out a pre-synaptic current spike is resistively split and then reamplified with an amplifying Josephson transmission line (JTL), which can be seen as the increasing $I_c$ values of the junctions in fig. 3a.  We have run simulations of this scheme at up to a direct division of 1-to-10.  This structure is also nestable, meaning that using 1-to-10 splitters one could get to a fan-out of 1-to-1000 in three layers.  It should be easier to modify this structure anywhere between a fan-out of 2 and a fan-out of 10 for any single stage, which could reduce the potential overhead if a non-power-of-two fan-out is desired.  However, a downside of this design is that when nesting 1-to-10 fan-out splitters, the JTL amplifier required at least 4 stages meaning that the final junction count overhead is ~ $4N_{FO}$ where $N_{FO}$ is the final fan-out number.  More precisely, in the case where the reamplification takes an M-junction JTL, the junction count is $N_{FO} = \sum_{i=1}^{k} M \cdot J^i$, where k is the number of stages and J is the single stage fan-out (10 in this example.)  Figure 3b shows the output of the last stage junction at each of the three stages in a simulated 1-to-1000 fan-out.  We note that the final stage is terminated with a resistor and inductor, which is why its pulse shape is slightly different from the other two stages as they are driving a 10 Ω resistor divide plus the first stage of the JTL re-amplifier. In this example we again feed an initial DC to SFQ converter with a 25 GHz triangle wave, its output is fed into the input node marked in Fig. 3a.

**IV Fan-in**

In this section we will discuss two methods for fan-in that can be used in neuromorphic SFQ circuits. Unlike the fan-out, we envision the fan-in to be an analog operation. Figure 4a shows the basic schematic for a flux-based fan-in. This circuit is designed for the input on each of the lines ($I_{in\_1}$ through $I_{in\_N_{FI}}$) to have already been weighted by a synaptic circuit. With flux-based fan-in, excitatory or inhibitory connections can be made using positive or negative inductive coupling. The summation loop has $N_{FI}$ small mutual inductors, two JJs and one more inductor to adjust the behavior of the loop, as well as a bias current, which can adjust the threshold.

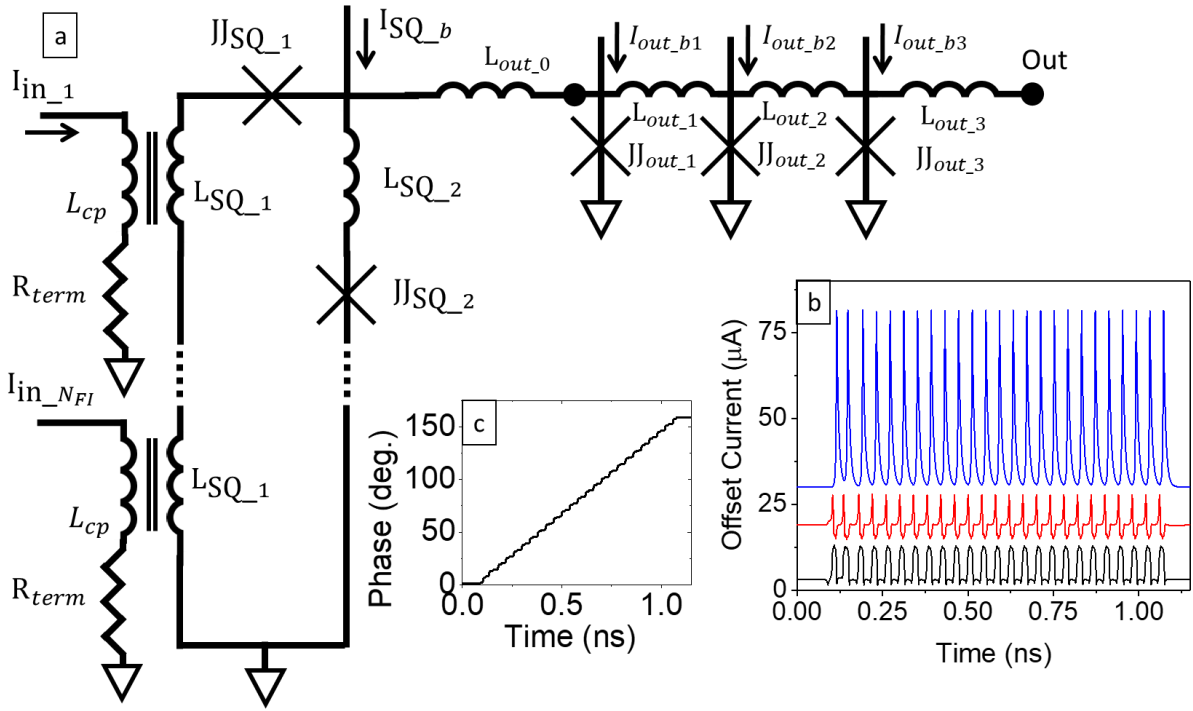

Figure 4a) Schematic of flux-based fan-in loop and JTL re-amplification. 4b) fan-in loop response to a single full weight spiking event for a fan-in of 128-to-1. Black trace shows the current across one of the coupling inductors $L_{SQ\_1}$, the red trace shows the current across $L_{out\_0}$, and the blue trace show the current after reamplification at $L_{out\_3}$. 4c) Shows the phase across $JJ_{SQ\_2}$ as it emits SFQ pulses into the re-amplification circuit. The data shown in 4b are from a simulation with the following circuit parameter; $JJ_{SQ\_1} = JJ_{2\_SQ}$ with $I_c$ = 12 µA, $L_{cp\_1}$ = 1 pH, $L_{SQ\_2}$ = 1 pH, $I_{b\_SQ}$ = 17 µA, $L_{out\_0}$ = 86pH, $JJ_{out\_1}$ has $I_c$ = 12 µA, $JJ_{out\_2}$ has $I_c$ = 24 µA, $JJ_{out\_3}$ has $I_c$ = 48 µA, $I_{b1}$ = 10 µA, and $I_{b2}$ = 20 µA, $I_{b3}$ = 40 µA, $L_{out\_1}$ = 86 pH, $L_{out\_2}$ = 43 pH, $L_{out\_3}$ = 22 pH.

While both fan-out circuits were digital and nested to achieve arbitrary levels of fan-out, the fan-in circuit is inherently analog and therefore cannot be nested in the same way. This means that the scaling of these circuits will be under further constraints, the details of which are discussed below. In the simulations presented here, we take a worst-case scenario of trying to drive the summation loop that is the post synaptic JJ neuron above threshold with a single incoming SFQ pulse. For small fan-in sizes this is relatively easy, and we show here that even under these "worst-case" conditions we are able to successfully drive a fan-in of 128-to-1. In order to drive the summation loop above threshold with a single input we had to decrease the critical current of the JJs in the post synaptic JJ neuron. If one wants to extend the network further than a single layer, we then need to reamplify the signal back to the original level so that it can be fanned-out into the next layer of the network. We include this re-amplification in our simulated circuits as shown in fig. 4a, with the amplifying JTL between $L_{SQ\_1}$ and the output.

Figure 4b shows the results of the simulation of a 128-to-1 fan-in circuit operating at 25 GHz under the single active input condition. In this simulation, we started with the fan-out circuit described in Fig. 2. However, only 1 of the final 128 junctions was connected back to the input pulses coming from the DC to SFQ input. The rest of the splitter network was terminated to ground on the input to ensure that there were no spikes coming from these JJs while maintaining the impedance of the network to simulate any potential cross talk effects. In our simulation one full-weight pulse is then defined by the final stage of a JTL amplifier after the fan-out with $I_c$ = 500 μA.

The traces in Fig. 4b shows the current pulse through one of the coupling inductors ($L_{SQ\_1}$ in Fig. 4a) in black, the output current of the summation loop through $L_{out\_0}$ is shown in red, and the output current through $L_{out\_3}$ after reamplification is shown in blue. Figure 4c shows the phase response of $JJ_{SQ\_2}$, exhibiting 2π phase steps for each output pulse of the post-synaptic JJ neuron, which confirms the desired SFQ type operation.

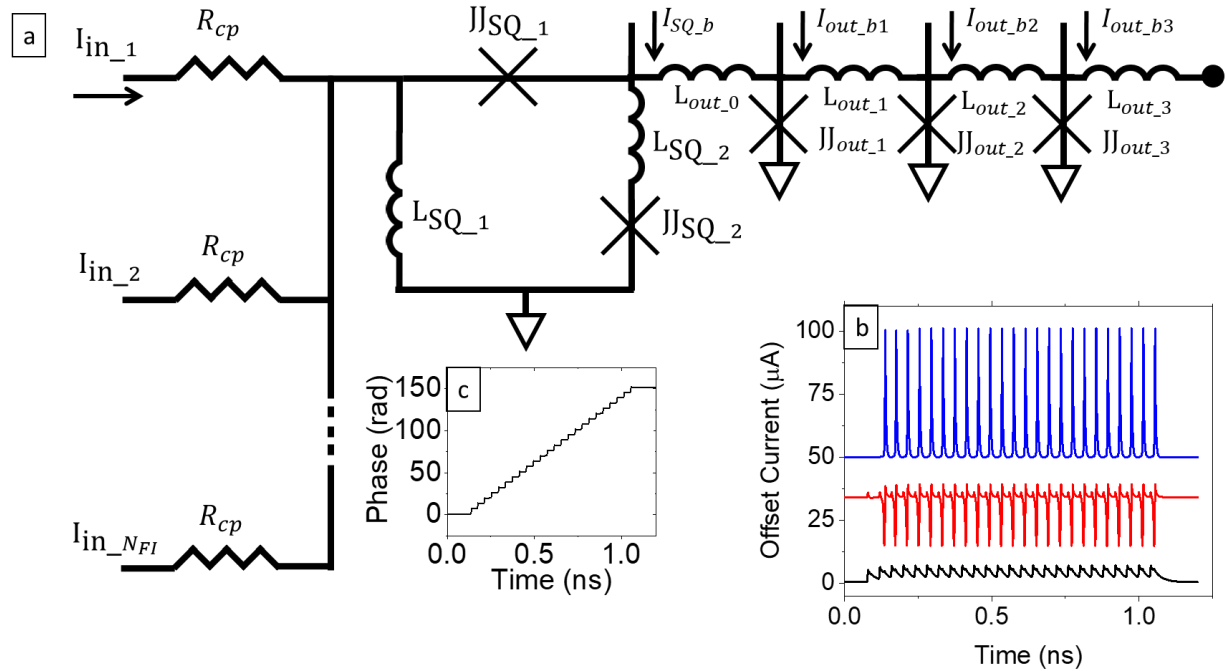

Figure 5a) Schematic of current-based fan-in with JTL re-amplification. 4b) fan-in response to a single full weight spiking event for a fan-in of 10-to-1. Black trace shows the current across $L_{SQ\_1}$, the red trace shows the current across the output of the fan-in / post-synaptic neuron $L_{out\_0}$, and the blue trace show the current after reamplification at $L_{out\_3}$. 4c) Shows the phase across $JJ_{SQ\_2}$ as it emits SFQ pulses into the re-amplification circuit. The data shown in 5b are from a simulation with the following circuit parameter; $R_{cp}$ = 10 Ohms, $JJ_{SQ\_1}$ and $JJ_{SQ\_2}$ have $I_{c1}$ = $I_{c2}$ = 20 μA, $I_{SQ\_b}$ = 40 μA, $L_{SQ\_1}$ = $L_{SQ\_2}$ = 40 pH, $L_{out\_0}$ = 40 pH, $L_{out\_1}$ = 30 pH, $L_{out\_2}$ = 20 pH, $L_{out\_3}$ = 17 pH, $JJ_{out\_1}$ has $I_c$ = 30 μA, $JJ_{out\_2}$ has $I_c$ = 40 μA, $JJ_{out\_3}$ has $I_c$ = 50 μA, $I_{out\_b1}$ = 26 μA, $I_{out\_b2}$ = 34 μA, $I_{out\_b3}$ = 42 μA.

In addition to the flux-based fan-in just described, we also investigated current-based fan-in. Figure 5a shows the basic schematic for a current-based fan-in. This circuit is designed for the input on each of the lines ($I_{in\_1}$ through $I_{in\_N_{FI}}$) to have already been weighted by a synaptic circuit. We again run simulations for the "worst-case" scenario of trying to drive the post-synaptic JJ neuron with a single input pulse. Figure 5b shows the results of simulations of the current-based fan-in of 10-to-1 operating at 25 GHz. The current pulse through $L_{SQ\_1}$ in the summation loop is shown in black, output current of the summation loop through $L_{out\_0}$ is shown in red, and the output current through $L_{out\_3}$ after reamplification is shown in blue. Figure 5c shows the phase evolution of $JJ_{SQ\_2}$, which demonstrates both 2π phase steps that accompany the SFQ pulsing of the post-synaptic JJ neuron and the desired SFQ operation of this part of the circuit. Note, $I_{in\_1}$ was taken as an output spike from a flux-based fan-out of 1-to-10 and was then amplified with a JTL amplifier with a final stage of 200 µA. All other inputs were grounded so as not to cause any additional input current, but to properly model the loss of current running back into the inactive inputs, which we will refer to as the cross-talk.

**V Constraints and limits for large networks**

For larger networks, the fan-out methods presented in Section III will be limited only by area, junction count, power, and delay. There is no fundamental limit to how many signals can be fanned out with either the splitter method or the current method. The scaling of the junction count with fan-out for each method is given in fig. 6. The splitter method seems like the better choice but depending on other design constraints it is possible that current method might still be viable.

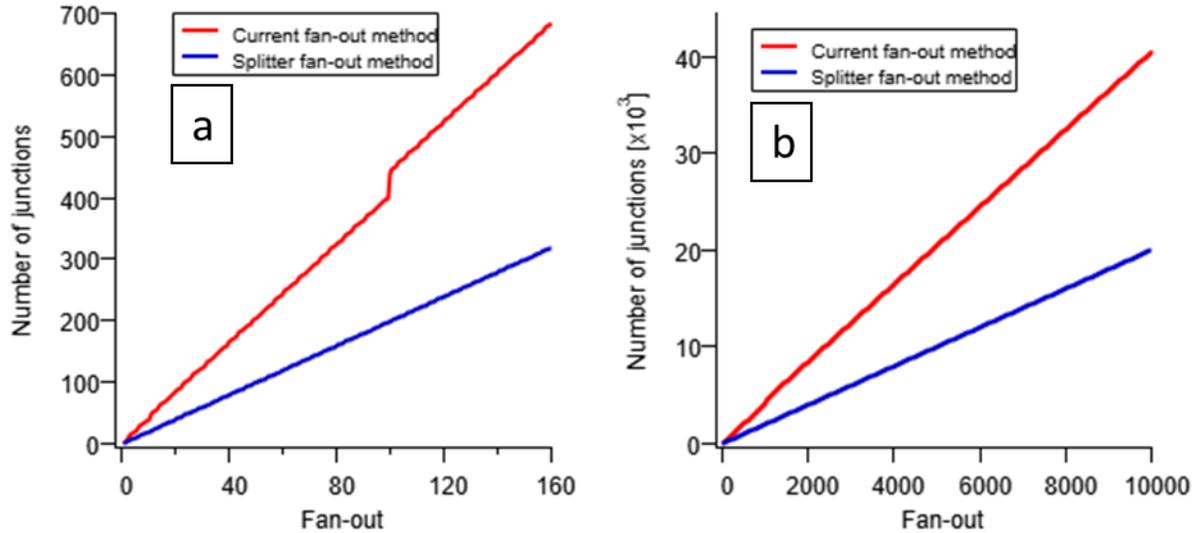

*Figure 6a: Number of junctions needed for a given fan-out for a smaller network and 6b) a larger, brain-sized network. The two different methods, current and splitter, are both shown. The jump in the current method at the powers of 10 occur because a new nested layer is needed.*

Depending on the architecture chosen for a superconducting SNN one may want to maintain a constant pulse timing. Figure 7a shows a block diagram schematic of binary splitters connected in a way that maintains a constant time delay at a single output line. In this example the timing is accomplished using constant wire path lengths, and junction counts for all outputs. If one were to relax the timing requirement or the single output line requirement, then the circuit size could be reduced. In the example shown in Fig. 7a the splitters were nested to three layers giving a fan-out is 1-to-8. One could also extend this structure arbitrarily deep depending on the desired level of fan-out. In this case, the size of the fan-out circuit will be determined primarily by two factors. The size in the vertical direction (as illustrated in Fig. 7a) is set by the minimum pitch of the JJs, and the size in the horizontal direction is set by the required inductance.

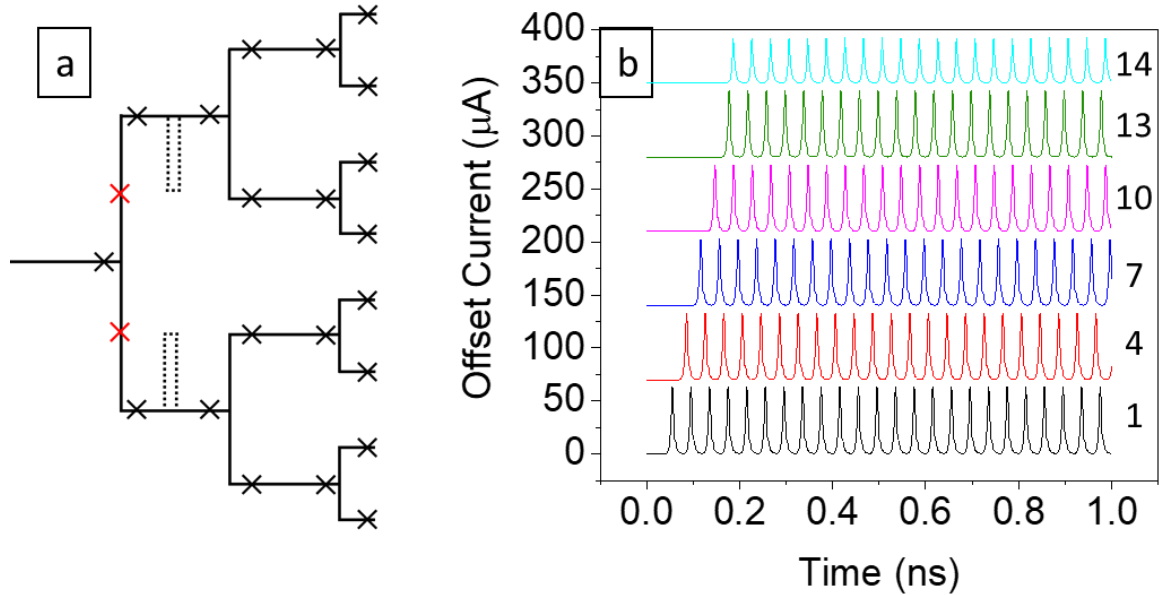

*Figure 7a) Basic block diagram schematic (not to scale) of a nested binary fan-out structure that maintains a constant time delay at the output. Note, bias currents and ground connections are not shown for clarity. The junctions shown in red are optional repeater junctions that may be required for larger fan-out implementations. The dashed lines illustrate potential meander lines that could be used to reduce the horizontal dimension. 2b) Example output current pulses, with layers indicated on the right, in a 1-to-16,384 fan-out circuit of nested binary splitters (pulses are offset for clarity with the first stage at the bottom and the last stage at the top). The data shown in 2b were from a simulation with the same unit cell parameters used in Fig. 2.*

Using this nominal two wiring layer layout, we estimate that a 1-to-128 fan-out circuit would occupy roughly 175 µm x 200 µm. To arrive at this estimated area, the following assumptions were used: The minimum JJ diameter is 500 nm, and the separation between JJ edges is 1 µm. This yields a 1.5 µm pitch for the JJ's in the final output line, which result in a maximum vertical length of ~200 µm for a fan-out of 1-to-128. To estimate the space required in the corresponding horizontal direction we assumed that the inductance of the strip-lines used for wiring can be varied between roughly 0.05 pH/µm and 0.9 pH/µm. Using the structure shown in Fig. 7a with the wiring as the source of inductance we estimate that the horizontal length of a 1-to-128 fan-out circuit would be ~ 175 µm. Reduction in the horizontal length could be achieved with the use of high inductance layers or by adding meanders in the wiring, an example of which is shown in Fig. 7a with dashed lines. Reduction in the vertical length could be achieved by reducing the JJ diameter and/or by reducing the space between JJs.

If space and time delay are permitted, then the binary splitters can be nested much deeper, including to levels similar to the human brain. Figure 7b shows example output current spikes from a fan-out circuit using the same basic parameters as outlined in Fig. 2 but extending the circuit to 14 layers deep resulting in a fan-out of 1-to-16,384. This 1-to-16,384 fan-out circuit would occupy an area of roughly 350 µm x 25mm using the same assumption discussed in Fig. 7a. This circuit is quite large, and one would hope that further scaling of the JJ technology will help to reduce this size in the future.

There is one additional constraint that must be considered as the fan-out is nested to larger circuits, which is the inclusion of repeater junctions (or JTLs), shown in red in Fig. 7a. These are needed once the impedance of the wire becomes too large for the JJs to drive within the SFQ regime and will minimally increase the junction count. In the 1-to-16,384 fan-out example, the longest wire (following the nominal layout of Fig. 7a) is roughly 12,300 µm. This would require roughly 122 additional junctions spread through the first 5 layers of the fanout, if we assume wiring with 0.05 pH/µm and a desired inductance of 20 pH. Compared to the total junctions count of roughly 3N (49,152) this is a negligible design constraint.

In addition to the flux-based splitter fan-out, one can also nest the current-based fan-out to an arbitrarily large size. The footprint of this type of fan-out will be similar in size to the splitter tree if we are constrained to a vertical line output with uniform time delay wiring. The rough size of a fan-out of 1-to-1000 (three layers deep) with this method is 1500 µm in the vertical direction and 250 µm in the horizontal direction. These dimensions were limited by the 1.5 µm pitch in the vertical direction, and by the 0.9 pH/µm maximum inductance in the horizontal direction.

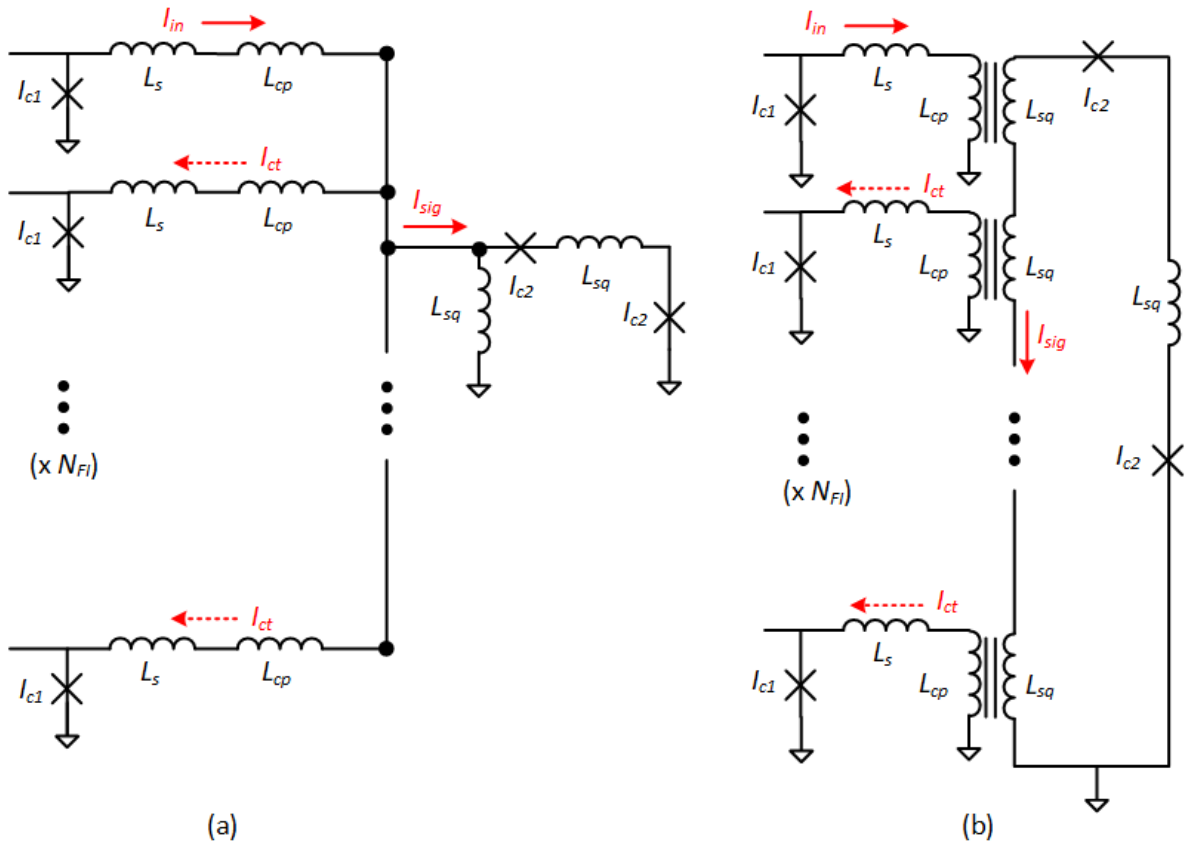

*Figure 8 : Simplified schematics for fan-in. (a) In the current method, the synaptic currents come to a common node and are sent through a two-junction loop. (b) In the flux method, the synaptic currents are inductively coupled into a common loop with two junctions. The input current ($I_{in}$), signal current ($I_{sig}$) and crosstalk current ($I_{ct}$) are indicated in red.*

Next, we discuss the limits to the two methods of fan-in, current and flux, shown earlier in Section IV. Figure 8 shows a generalized schematic of each method where we have tried to simplify each circuit to contain only the essential parts that affect scaling to large $N_{FI}$. Figure 8a shows the current method of fan-in while Fig. 8b shows the flux method. In both cases, the input to the fan-in coming down each synapse is an SFQ pulse with current $I_{in}$, generated by a junction with critical current $I_{c1}$ and a total series inductance ($L_s + L_{cp}$). In the current method, this current that flows into a summing node and then into an output superconducting quantum interference device (SQUID), which has an inductance $L_{sq}$ and a critical current of $I_{c2}$. In the flux method, the current $I_{in}$ flows through the coupling inductance $L_{cp}$ which is magnetically coupled to the inductance $L_{sq}$. This causes a flux in the output loop which has two junctions with critical currents equal to $I_{c2}$.

With these simplified circuits we can easily derive the value of the signal current $I_{sig}$ that flows into the output loop in both cases. In the case of the current method, we use the equation for current division. We first calculate the total parallel inductance $L_{||}$ from the summing node to ground:

$$L_{||} = \left(\frac{1}{L_{sq}} + \frac{1}{L_{cp}+L_s} + \frac{1}{L_{cp}+L_s} + \cdots\right)^{-1} = \left(\frac{1}{L_{sq}} + \frac{N_{FI}-1}{L_{cp}+L_s}\right)^{-1}, \tag{1}$$

where the term $1/(L_{cp} + L_s)$ appears ($N_{FI}$-1) times, once for each of the remaining fan-in paths aside from the one generating the pulse. For now, we have assumed that $L_{sq}$ is the inductance going to ground for the signal path, ignoring the higher-impedance path through the junctions; we will correct this later. The signal current is then simply $I_{in}\left(\frac{L_{||}}{L_{sq}}\right)$. This leads to:

$$I_{sig} = \frac{L_{cp}+L_s}{(N_{FI}-1)L_{sq}+(L_{cp}+L_s)}I_{in}. \tag{2}$$

Meanwhile, for the flux method we use the equations of mutual inductance. With a current of $I_{in}$ flowing through $L_{cp}$, the flux coupled to the output loop is $MI_{in}$, where $M = k\sqrt{L_{cp}L_{sq}}$ and $k$ is the coupling constant. This then results in a signal current of

$$I_{sig} = \frac{M}{L_{tot}}I_{in}, \tag{3}$$

where $L_{tot}$ is the total inductance of the large summing loop. Later we will relate $L_{tot}$ to $N_{FI}$ and $L_{sq}$.

Looking at Eqs. (2) and (3), we can make a few assumptions about some of the terms. First, we note that the current $I_{in}$ which appears in Eqs. (2) and (3) is the result of an SFQ pulse, and therefore has a total quantized flux of $\Phi_0$ = 2.07x10$^{-15}$ Vs. Since it is coupled through an inductance of ($L_{cp}$ + $L_s$), we can write

$$I_{in} = \frac{\Phi_0}{L_{cp}+L_s}. \tag{4}$$

In addition, since the both the input loop ($I_{c1} - L_s+L_{cp}$) and the output loop ($I_{c2} - L_{sq}$) each need to generate an SFQ pulse, there is a constraint on their inductance parameter $\beta_L$. Loops capable of generating SFQ pulses need to have a value of $\beta_L$ in the range $\beta_L \approx 1 - 10$.[23] Loops with $\beta_L < 1$ are non-hysteretic and incapable of generating an SFQ pulse; loops with $\beta_L > 10$ can store more than one flux quantum and are subject to random phase slips. We define $\beta_{L1}$ and $\beta_{L2}$ as follows:

$$\beta_{L1} = \frac{2\pi I_{c1}(L_s+L_{cp})}{\Phi_0} \tag{5}$$

$$\beta_{L2} = \frac{2\pi I_{c2} L_{sq}}{\Phi_0}. \tag{6}$$

In our designs both of these parameters ended up being between 1.5 and 4 and close to each other. (Note that $\beta_{L2}$ in eq. (6) is defined for the current fan-in case, in the flux case there would be $N_{FI}$ SQUID inductors.) Finally, we refer the signal current to the critical current $I_{c2}$, which is the junction that must release an SFQ pulse when the sum of the fan-in signals exceed threshold. Ideally, $I_{sig}$ should be a significant fraction of $I_{c2}$, in the neighborhood of at least 20%. Although smaller signal currents (1%-10%) could still potentially cause pulses in the output loop, this would require an external bias current very close to $I_{c2}$. In that case fabrication margins will become a significant concern. We thus use 20% as a rough rule of thumb.

We now apply these three assumptions into the equations for the signal and then compare the expression for the two methods. Looking first at the current, we find:

$$\frac{I_{sig}}{I_{c2}} = \frac{1}{(N_{FI}-1)\frac{\beta_{L2}}{2\pi}+\left(\frac{I_{c2}}{I_{c1}}\right)\frac{\beta_{L1}}{2\pi}} \approx \frac{1}{N_{FI}}\left(\frac{2\pi}{\beta_{L2}}\right), \tag{7}$$

where we have assumed that $N_{FI} \gg 1$ and $I_{c1} \gg I_{c2}$ (chosen to maximize the signal – see below), thus ignoring the second term in the denominator. For the case of flux, we make the simple assumption that $L_{tot}$ in equation (3) is equal to $N_{FI}L_{sq}$, again assuming that $N_{FI} \gg 1$ and also ignoring the slight modification to the impedance of each of the $L_{sq}$ inductors due to the coupling back to the input. Then we obtain:

$$\frac{I_{sig}}{I_{c2}} \approx \frac{k}{N_{FI}}\left(\frac{2\pi}{\beta_{L1}}\right)\left(\frac{I_{c1}}{I_{c2}}\right)\sqrt{\frac{L_{cp}}{L_{sq}}}. \tag{8}$$

In our case $L_{cp} = L_{sq}$, so the term in the square root is 1. Note then the similarities between Eqs. (7) and (8). The one major difference is the factor of ($I_{c1}/I_{c2}$) in the flux case. (There is also the factor of $k$, but this can be of order one.) This allows scaling to a much larger fan-in in the flux case, as one can make this ratio of critical currents ($I_{c1}/I_{c2}$) a factor of 50 or even higher. Thus, the flux method scales to large $N_{FI}$ better than the current method. This is a major result of our paper. Besides the comparison between Eqs. (7) and (8), we will also show this directly with simulations.

The expressions for the signal current derived in Eqs. (2) and (3) along with their simplified form in (7) and (8) can be verified by circuit simulations. We built a WR-SPICE model of the circuits in Fig. 8, choosing parameters that were similar to those in the simulations shown in Section IV. In both the current and the flux case we chose $I_{c1}$ = 1000 μA. This allowed us to decrease the size of $L_{cp}$ and $L_s$ while keeping $\beta_{L1}$ roughly constant, via equation (5). Meanwhile, the smaller value of ($L_{cp} + L_s$) helped maximize the size of the current $I_{in}$, via equation (4). The value of $I_{c2}$, on the other hand, was kept as small as possible to account for a small signal current. We used both a value of $I_{c2}$ = 20 μA, which is safe from the effects of thermal fluctuations at 4 K, and a value of 6 μA, which is very aggressive. Our values of $L_s$ and $L_{cp}$ were 0.8 pH and 1 pH in the current case and both equal to 1 pH in the flux case. The value of $L_{sq}$ is different for the flux and the current case and is described below.

After setting these parameters we varied the fan-in and noted how the signal current changed. Figures 9 and 10 show the cases for current and flux, respectively, where we plot the ($I_{sig}/I_{c2}$) as a function of $N_{FI}$. The markers show the results from circuit simulations while the solid and dashed lines show the results from the equations. The dashed line at a value of ($I_{sig}/I_{c2}$) = 0.2 represents our 20% criterion described above.

Looking first at the current, we used equation (7) for the fitting, but with the value of $L_{sq}$ modified to include the extra path to ground in parallel. This path included the additional inductor $L_{sq}$ and the Josephson inductance of the two junctions with $I_{c2}$. This parallel path has an inductance of about $2L_{sq}$, so the total equivalent inductance to ground is then $2/3\ L_{sq}$. Thus, we replaced $L_{sq}$ with $2/3\ L_{sq}$ to calculate $\beta_{L2}$ in equation (6), which is then used in equation (7) to calculate the signal current. Figure 9 shows the simulated data and fits for three cases: (i) $L_{sq}$ = 40 pH and $I_{c2}$ = 20 µA; (ii) $L_{sq}$ = 133 pH and $I_{c2}$ = 6 µA; (iii) $L_{sq}$ = 133 pH and $I_{c2}$ = 20 µA. The equations predict the reduction in signal very well. The maximum fan-in, when the signal drops to 20% as denoted by the black dashed line, is about $N_{FI}$ = 20. Note there is really no way to improve this; going lower in $I_{c2}$ means having to go higher in $L_{sq}$, which does not increase the signal. Thus, the red and green curves represent about the best performance possible.

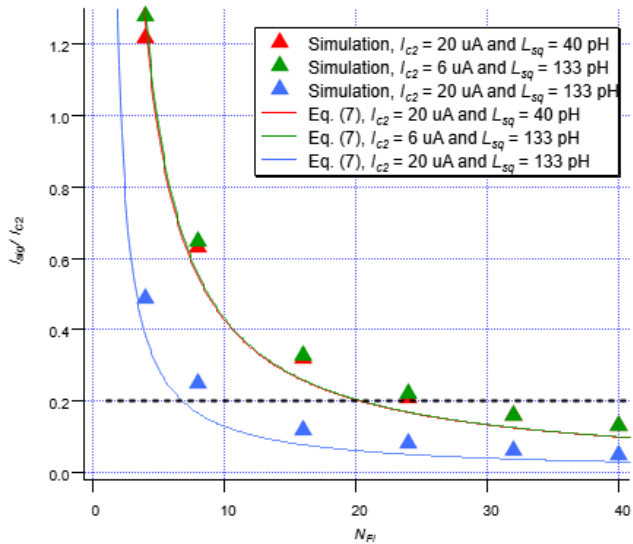

Figure 9: Fan-in scaling for the current method. The symbols show the simulations while the lines show the fits from equation (7). Red is for a value of $I_{c2}$ = 20 uA and $L_{sq}$ = 40 pH, green is for $I_{c2}$= 6 uA and $L_{sq}$ = 133 pH, and blue is for $I_{c2}$ = 20 uA and $L_{sq}$ = 133 pH. Other parameters in the simulation include $I_{c1}$ = 1000 uA and Ls = $L_{cp}$ = 1 pH. The black dotted line shows the criterion of a 20% signal. The red and green cases have the same value of $\beta_{L2}$, so according to equation (7) they should have the same signal, which they do.

Next, we look at fitting the case for flux fan-in. We use $L_{sq}$ = 1 pH, trying to keep it as small as possible since there are so many of them. Figure 10 shows the simulated data for the two values of $I_{c2}$, 20 µA and 6 µA. The dotted lines show the fit from equation (8), which assumes that the total inductance in the large loop is $N_{FI}L_{sq}$. These fit the data well at large $N_{FI}$, but not at small $N_{FI}$. As with the current method, we need to account for the other components in the loop, the two junctions and the remaining $L_{sq}$ inductor; we call this extra inductance $L_{ex}$, which simply adds to $N_{FI}L_{sq}$. In addition, since each of the $L_{sq}$ inductors couple a small current back to the input loop (the cross-talk current, which will be discussed below), their impedance is reduced below the value of $L_{sq}$. The reduced value $L'_{sq}$ is given by:

$$L'_{sq} = L_{sq} - \frac{M^2}{L_{cp}+L_s}, \tag{9}$$

which can be derived from the circuit model of a transformer. Incorporating both of these corrections, we can then rewrite equation (8) as

$$\frac{I_{sig}}{I_{c2}} = \frac{k}{N}\left(\frac{I_{c1}}{I_{c2}}\right)\left(\frac{2\pi}{\beta_{L1}}\right)\sqrt{\frac{L_{cp}}{L_{sq}}}\left[\frac{1}{1-k^2 L_{cp}/(L_{cp}+L_s)+L_{ex}/(N_{FI}L_{sq})}\right]. \tag{10}$$

The corrective term is in square brackets. In Fig. 10 the solid lines show the fits using equation (10), showing better agreement at smaller $N_{FI}$. At larger values of $N_{FI}$ the corrective term gets close to one, thus the curves become roughly the same. The maximum fan-in for flux is over 100 with the 20 µA case and over 300 in the 6 µA case. Note these are much larger than the fan-in for current, which confirms our conclusion that flux scales better than current.

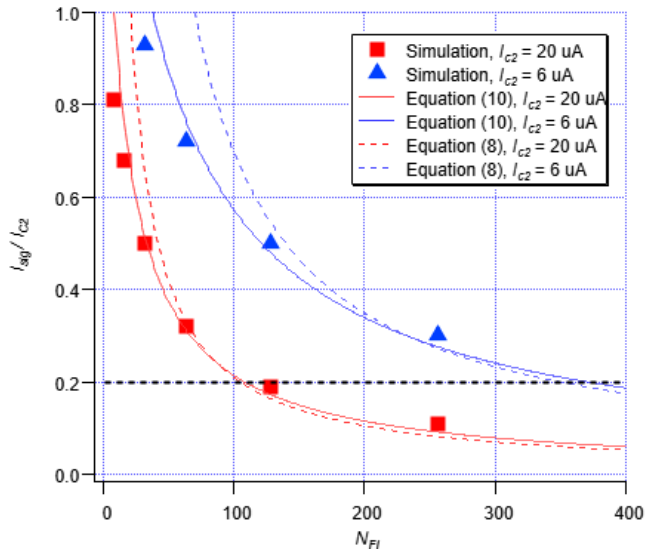

Figure 10: Fan-in scaling for the flux method. The symbols show the simulations while the lines show the fits from Eqs. (8) and (10). Red is for a value of $I_{c2}$ = 20 µA, while blue is for $I_{c2}$ = 6 µA. Other parameters in the simulation include Ic1 = 1000 µA, $L_{sq}$ = $L_{cp}$ = 1 pH, $L_{ex}$ = 14 pH (red case), $L_{ex}$ = 38 pH (blue case), and $L_s$ = 0.8 pH. The black dotted line shows the criterion of a 20% signal.

Finally, we look at the crosstalk current ($I_{ct}$) which flows back down synapses, shown in Fig. 8. For the current method, $I_{ct}$ is given by:

$$I_{ct} = \left(\frac{L_{\|}}{L_{cp}+L_s}\right)I_{in} = \frac{L_{sq}}{(L_{cp}+L_s)+(N_{FI}-1)L_{sq}}I_{in}. \tag{11}$$

Using our numbers for the case where $N_{FI}$ = 16, $L_{sq}$ = 40 pH and $I_{c2}$ = 20 µA, we find a crosstalk current of 33.2 µA using equation (11), which compares well to our simulated value of 32.6 µA. This is actually much larger than the signal current for those parameters (6.4 µA). Note that the crosstalk current scales approximately as (1/$N_{FI}$), so the inability to increase the fan-in keeps the crosstalk large.

In the flux case, the crosstalk is given by:

$$I_{ct} = \frac{M}{L_s+L_{J1}} I_{sig} \approx \frac{M^2}{N_{FI} L_{sq}(L_s+L_{J1})} I_{in}, \qquad (12)$$

where $L_{J1}$ is the Josephson inductance of the input junction with $I_{c1}$ and we have approximated $N_{FI} \gg 1$. Using our numbers for the case where $N_{FI}$ = 128 and $I_{c2}$ = 20 µA we find a crosstalk current of 1.3 µA using equation (12), the same as what we find in our simulation and much smaller than in the current method. Here we also find that the crosstalk current scales approximately as (1/$N_{FI}$), so the ability to go to large $N_{FI}$ helps reduce the crosstalk. The size of the crosstalk for flux can thus be made much smaller than for current, which once again confirms our conclusion that flux is indeed the better choice for fan-in.

**VI Discussion**

In this paper we have explored the problem of fan-out and fan-in in superconducting neuromorphic circuits. The fan-out is a digital process where the action potential is repeated and copied multiple times. Both flux and current methods give fan-out schemes that accomplish this successfully. Both can be scaled indefinitely to large networks, with a proportional cost in chip area and power dissipation. Since power dissipation scales with the junction count, and each junction dissipating roughly $I_c \Phi_0$ energy per SFQ pulse, a rough estimate of the power dissipated for a 1-to-128 flux based fan-out circuit using the $I_c$ values given above is 44 aJ. Pulse timing can be preserved throughout subsequent layers of fan-out; if the constraint on timing is lifted, then the cost in area can be reduced. The flux method has a better scaling with junction number, but current fan-out might still be used, depending on the application.

Fan-in is an analog process where the weighted synaptic signals are summed and compared to a threshold. Both a current method and flux method were proposed. In scaling to large networks, the flux method appears to be superior to the current method, both in preserving a larger signal current and in keeping the cross-talk to a minimum. For small networks it is possible that the current method could still be used.

In our study we have largely ignored the problem of synaptic weighting; that is covered in other work.[5] However, the architectural assumptions that we have made in this paper are consistent with typical SFQ weighting schemes that have been proposed in the literature. In addition, in our study of the fan-in we have been careful to consider the worst-case scenario for the signal, insisting that a single synaptic current be capable of invoking an action potential. In real networks there will most likely be signals from many synapses simultaneously, increasing the total signal beyond the estimates in this paper.

Looking forward, there are many possible advantages to a future superconducting neuromorphic processor: (i) the ability to use superconducting multi-chip modules (MCMs)[24] to connect many chips together with no cost in dissipation; (ii) the availability[25] of low-power superconducting digital electronics to provide interfacing, multiplexing and readout between analog units; (iii) the ability of the JJ neuron to attain biological realism[1] with only two junctions; and (iv) possible optical interface to truly expand to brain-sized networks. In this work, we demonstrate that both a fan-in and fan-out of order 100 are completely reasonable for existing superconducting fabrication. Using our area estimates, a network of 128 fully-connected neurons would easily fit on a 5 mm x 5 mm chip and draw minimal power. Thus, we conclude that the fan-in and fan-out are not limiting factors for the future of superconducting neuromorphic computing.

**Data Availability Statement:**

The data that supports the findings of this study are available within the article [and its supplementary material].

**References:**

[1] P. Crotty, D. Schult, and K. Segall, Phys. Rev. E **82**, 011914 (2010).
[2] M.L. Schneider, C.A. Donnelly, S.E. Russek, B. Baek, M.R. Pufall, P.F. Hopkins, P.D. Dresselhaus, S.P. Benz, and W.H. Rippard, Sci. Adv. **4**, e1701329 (2018).
[3] K. Segall, M. LeGro, S. Kaplan, O. Svitelskiy, S. Khadka, P. Crotty, and D. Schult, Phys. Rev. E **95**, 032220 (2017).
[4] K. Segall, S. Guo, P. Crotty, D. Schult, and M. Miller, Phys. B-Condens. Matter **455**, 71 (2014).
[5] J.M. Shainline, IEEE J. Sel. Top. Quantum Electron. **26**, 1 (2020).
[6] J.M. Shainline, S.M. Buckley, A.N. McCaughan, J.T. Chiles, A.J. Salim, M. Castellanos-Beltran, C.A. Donnelly, M.L. Schneider, R.P. Mirin, and S.W. Nam, J. Appl. Phys. **126**, 044902 (2019).
[7] F. Chiarello, P. Carelli, M.G. Castellano, and G. Torrioli, Supercond. Sci. Technol. **26**, 125009 (2013).
[8] T. Onomi and K. Nakajima, J. Phys. Conf. Ser. **507**, 042029 (2014).
[9] I.I. Soloviev, A.E. Schegolev, N.V. Klenov, S.V. Bakurskiy, M.Yu. Kupriyanov, M.V. Tereshonok, A.V. Shadrin, V.S. Stolyarov, and A.A. Golubov, J. Appl. Phys. **124**, 152113 (2018).
[10] R. Cheng, U.S. Goteti, and M.C. Hamilton, J. Appl. Phys. **124**, 152126 (2018).
[11] Y. Yamanashi, K. Umeda, and N. Yoshikawa, IEEE Trans. Appl. Supercond. **23**, 1701004 (2013).
[12] T. Hirose, T. Asai, and Y. Amemiya, Phys. C Supercond. **463–465**, (2007).
[13] W. Maass, Neural Netw. **10**, 1659 (1997).
[14] W. Gerstner, H. Sprekeler, and G. Deco, Science **338**, 60 (2012).
[15] D.S. Holmes, A.M. Kadin, and M.W. Johnson, Computer **48**, 34 (2015).
[16] O.A. Mukhanov, IEEE Trans. Appl. Supercond. **21**, 760 (2011).
[17] P. Bunyk, K. Likharev, and D. Zinoviev, Int. J. High Speed Electron. Syst. **11**, 257 (2001).
[18] M.L. Schneider, C.A. Donnelly, S.E. Russek, B. Baek, M.R. Pufall, P.F. Hopkins, and W.H. Rippard, in *2017 IEEE Int. Conf. Rebooting Comput. ICRC* (2017), pp. 1–4.
[19] N. Katam, A. Shafaei, and M. Pedram, in *2017 22nd Asia S. Pac. Des. Autom. Conf. ASP-DAC* (IEEE, Chiba, Japan, 2017), pp. 384–389.
[20] P. Dayan and L.F. Abbott, *Theoretical Neuroscience: Computational and Mathematical Modeling of Neural Systems* (Massachusetts Institute of Technology Press, Cambridge, Mass, 2001).
[21] P.A. Merolla, J.V. Arthur, R. Alvarez-Icaza, A.S. Cassidy, J. Sawada, F. Akopyan, B.L. Jackson, N. Imam, C. Guo, Y. Nakamura, B. Brezzo, I. Vo, S.K. Esser, R. Appuswamy, B. Taba, A. Amir, M.D. Flickner, W.P. Risk, R. Manohar, and D.S. Modha, Science **345**, 668 (2014).
[22] V. Sze, Y.-H. Chen, T.-J. Yang, and J.S. Emer, Synth. Lect. Comput. Archit. **15**, 1 (2020).
[23] T. Van Duzer and C.W. Turner, (1981).

[24] S. Narayana, V.K. Semenov, Y.A. Polyakov, V. Dotsenko, and S.K. Tolpygo, Supercond. Sci. Technol. **25**, 105012 (2012).

[25] R.N. Das, V. Bolkhovsky, S.K. Tolpygo, P. Gouker, L.M. Johnson, E.A. Dauler, and M.A. Gouker, in *2017 IEEE 67th Electron. Compon. Technol. Conf. ECTC* (2017), pp. 675–683.